\documentclass[a4paper,12pt]{article}

\usepackage{graphics, epsfig}

\begin{document}

\author{C. Barrab\`es\thanks{E-mail : barrabes@celfi.phys.univ-tours.fr}
and P.A. Hogan\thanks{E-mail : peter.hogan@ucd.ie}\\
\small Laboratoire de Math\'ematiques et Physique Th\'eorique\\
\small  CNRS/UMR 6083, Universit\'e F. Rabelais, 37200 TOURS,
France}

\title{Scattering of High Speed Particles in the Kerr Gravitational Field}
\date{}
\maketitle

\begin{abstract}
We calculate the angles of deflection of high speed particles
projected in an arbitrary direction into the Kerr gravitational
field. This is done by first calculating the light--like boost of
the Kerr gravitational field in an arbitrary direction and then
using this boosted gravitational field as an approximation to the
gravitational field experienced by a high speed particle. In the
rest frame of the Kerr source the angles of deflection experienced
by the high speed test particle can then easily be evaluated.

\end{abstract}
\thispagestyle{empty}
\newpage

\section{Introduction}\indent
The scattering of a high speed particle in the Kerr gravitational
field is examined using a novel viewpoint first introduced by the
authors in \cite{CQG}. In the rest frame of the particle the Kerr
field is distorted, as the speed of the particle relative to the
source of the Kerr field approaches the speed of light. In the
light--like limit the Kerr field becomes the field of an impulsive
gravitational wave. Thus from the point of view of a particle travelling
close to the speed of light the Kerr gravitational field can be
approximated by an impulsive gravitational wave field, and the 
path of the particle
calculated using the time--like geodesic equations of this
space--time. When the path is expressed in the rest--frame of the
Kerr source the angles of deflection can be deduced.
While in \cite{CQG} only the case of a particle projected
along a direction parallel to the equatorial plane was considered,
we here obtain the angles of deflection for an arbirary initial
direction of the particle. 
The results are most accurate for high speed
particles projected with large impact parameter. Under these
circumstances the leading terms in the calculated angles of
deflection agree with the known angles of deflection of photons in
the Kerr field. 

In order to examine the scattering of high speed
particles projected in an arbitrary direction into the Kerr field
from this point of view we first require the light--like boost of
the Kerr field in an arbitrary direction. This is given in section
2. The result is presented in a form which is well adapted to the
study of high-energy scattering processes such as the one considered
in this paper. Another application is the
classical production of $D$-dimensional
black holes in high-energy collisions \cite{Gid}.
In those scenarios the highly relativistic colliding particles are
modelled by plane--fronted impulsive gravitational waves, in order
to study the formation of a trapped surface after the collision of 
the two waves, and derive estimates of the mass of the resulting
black hole and of the classical cross--section for black hole production.

\setcounter{equation}{0}
\section{Boosted Kerr Field in an Arbitrary Direction}\indent
Our approach to analyzing the scattering of a high speed particle
in the Kerr gravitational field, or indeed in the gravitational
field of any isolated source, involves first calculating the
gravitational field of the isolated source as it appears to an
observer moving rectilinearly relative to the source with speed
approaching that of light (see \cite{CQG} and also \cite{book}).
To do this for the Kerr field we start with the Kerr line--element
in the Kerr--Schild form \cite{K}, \cite{KS}
\begin{equation}\label{1.1}
ds^2=ds_0^2+{2\,m\,\bar r^3\over \bar r^4+A^2\bar z^2}\,(\bar
k_{\mu}\,d\bar x^{\mu})^2=\bar g_{\mu\nu}\,d\bar x^{\mu}\,d\bar
x^{\nu}\ ,\end{equation} with \begin{eqnarray}\label{1.2}
ds_0^2&=&d\bar x^2+d\bar y^2+d\bar z^2-d\bar t^2=\bar\eta
_{\mu\nu}\,d\bar x^{\mu}\,d\bar x^{\nu} ,\\
\bar k_{\mu}\,d\bar x^{\mu}&=&{\bar z\over\bar r}\,d\bar z+\left
({\bar r\,\bar x+A\,\bar y\over \bar r^2+A^2}\right )\,d\bar
x+\left ({\bar r\,\bar y-A\,\bar x\over \bar r^2+A^2}\right
)\,d\bar y-d\bar t\ .\end{eqnarray}In (\ref{1.1}) $m$ is the mass
and ${\bf J}=(0, 0, m\,A)$ is the angular momentum of the source.
Also $\bar r$ is given in terms of $\bar x,\ \bar y\ ,\bar z$ by
\begin{equation}\label{1.4}\frac{\bar x^2+\bar y^2}{\bar r^2+A^2}+\frac{\bar z^2}{\bar
r^2}=1\ .\end{equation}The bars on coordinates are for convenience
and will be removed below. Our approach to calculating the
light--like boost of the gravitational fields of isolated sources
is centered on the Riemann tensor and thus we require a convenient
form of the Riemann tensor for the Kerr field. This is given, in
the barred coordinates, by the complex quantity
\begin{equation}\label{1.5}
{}^+\bar R_{\mu\nu\rho\sigma}=\bar R_{\mu\nu\rho\sigma}+i{}^*\bar
R_{\mu\nu\rho\sigma}\ ,\end{equation} where Greek indices take
values 1, 2, 3, 4 and the star indicates the dual of the Riemann
tensor. Thus ${}^*\bar R_{\mu\nu\rho\sigma}=\frac{1}{2}\bar\eta
_{\mu\nu\alpha\beta}\,\bar R^{\alpha\beta}{}_{\rho\sigma}$ with
$\bar\eta _{\mu\nu\alpha\beta}=\sqrt{-\bar g}\,\bar\epsilon
_{\mu\nu\alpha\beta}$ where $\bar g$ is the determinant of the
metric tensor in the barred coordinates and $\bar\epsilon
_{\mu\nu\alpha\beta}$ is the Levi--Civita permutation tensor in
the barred coordinates. For the Kerr space--time with
line--element (\ref{1.1}) above ${}^+\bar R_{\mu\nu\rho\sigma}$ is
given by \cite{PR}
\begin{equation}\label{1.6}
{}^+\bar R_{\mu\nu\rho\sigma}=-\frac{m\,\bar r^3}{(\bar
r^2+iA\,\bar z)^3}\,\left\{\bar g_{\mu\nu\rho\sigma}+i\bar\epsilon
_{\mu\nu\rho\sigma}+3\,\bar W_{\mu\nu}\,\bar
W_{\rho\sigma}\right\}\ ,\end{equation} where
\begin{equation}\label{1.7}\bar g_{\mu\nu\rho\sigma}=\bar g_{\mu\rho}\,\bar
g_{\nu\sigma}-\bar g_{\mu\sigma}\,\bar g_{\nu\rho}\
,\end{equation}and $\bar W_{\mu\nu}$ is given by the 2--form
\begin{eqnarray}\label{1.8}
{1\over 2}\,\bar W_{\mu\nu}\,d\bar x^{\mu}\wedge d\bar
x^{\nu}&=&\frac{\bar r}{\bar r^2+i\,A\,\bar z}\,[\bar x\,(d\bar
x\wedge d\bar t-i\,d\bar y\wedge d\bar z)+\bar y\,(d\bar y\wedge
d\bar t\nonumber\\&&-i\,d\bar z\wedge d\bar x)+(\bar z+iA)\,(d\bar
z\wedge d\bar t-i\,d\bar x\wedge d\bar y)]\ .\end{eqnarray}We
require the Kerr field when the angular momentum points in an
arbitrary direction in space and not in the positive $\bar
z$--direction as it does here. With $a, b, c$ real numbers such
that $\sqrt{a^2+b^2+c^2}=A$ we make the {\it rotation}
\begin{eqnarray}\label{1.9}
\bar x&\rightarrow & -{(a\,c^2+b^2A)\over A(b^2+c^2)}\,\bar
x-{b\,c\,(a-A)\over A\,(b^2+c^2)}\,\bar y+{c\over A}\,\bar z\
,\nonumber\\ \bar y&\rightarrow &-{b\,c\,(a-A)\over
A\,(b^2+c^2)}\,\bar x-{(a\,b^2+c^2A)\over A\,(b^2+c^2)}\,\bar
y+{b\over A}\,\bar z\ ,\\\bar z&\rightarrow &{c\over A}\,\bar
x+{b\over A}\,\bar y+{a\over A}\,\bar z\ .\nonumber\end{eqnarray}
Under this rotation ${\bf J}=(0, 0, mA)\rightarrow (mc, mb, ma)$
so that the angular momentum of the source points in an arbitrary
direction relative to the new $\bar x, \bar y, \bar z$ axes.
Substituting (\ref{1.9}) into (\ref{1.4}) we find that $\bar r$ is
given in terms of the new $\bar x, \bar y, \bar z$ by
\begin{equation}\label{1.10}
\bar x^2+\bar y^2+\bar z^2+\frac{(c\bar x+b\bar y+a\bar z)^2}{\bar
r^2}=\bar r^2+a^2+b^2+c^2\ .\end{equation}In addition the
following hold under (\ref{1.9}):
\begin{equation}\label{1.11}
d\bar x^2+d\bar y^2+d\bar z^2\rightarrow d\bar x^2+d\bar y^2+d\bar
z^2\ ,
\end{equation}
\begin{equation}\label{1.12}
\bar x\,d\bar x+\bar y\,d\bar y+(\bar z+i\,A)\,d\bar z\rightarrow
(\bar x+i\,c)\,d\bar x+(\bar y+i\,b)\,d\bar y+(\bar z+i\,a)\,d\bar
z\ ,\end{equation} \begin{equation}\label{1.13} \bar y\,d\bar
x-\bar x\,d\bar y\rightarrow \frac{(a\,\bar y-b\,\bar
z)}{A}\,d\bar x+\frac{(c\,\bar z-a\,\bar x)}{A}\,d\bar
y+\frac{(b\,\bar x-c\,\bar y)}{A}\,d\bar z\ ,\end{equation} and
\begin{eqnarray}\label{1.14}
\bar x\,d\bar y\wedge d\bar z + \bar y\,d\bar z\wedge d\bar x &+& (\bar
z+i\,A)\,d\bar x\wedge d\bar y\,\rightarrow (\bar x+i\,c)\,d\bar
y\wedge d\bar z \nonumber\\ &+&(\bar y+i\,b)\,d\bar z\wedge d\bar
x+(\bar z+i\,a)\,d\bar x\wedge d\bar y\
.\end{eqnarray}We can use these results to see the following:
Under (\ref{1.9}) the Kerr metric tensor retains the Kerr--Schild
form
\begin{equation}\label{1.15}
\bar g_{\mu\nu}=\bar\eta _{\mu\nu}+2\,H\,\bar k_{\mu}\,\bar
k_{\nu}\ ,\end{equation} with $\bar\eta _{\mu\nu}={\rm diag}(1, 1,
1, -1)$, \begin{equation}\label{1.16} H=\frac{m\,\bar r^3}{\bar
r^4+({\bf a}\cdot\bar {\bf x})^2}\ ,\end{equation} and
\begin{equation}\label{1.16'}
\bar k_{\mu}\,d\bar x^{\mu}=\frac{({\bf a}\cdot \bar {\bf
x})\,({\bf a}\cdot d\bar {\bf x})}{\bar r\,(\bar r^2+|{\bf
a}|^2)}+\frac{\bar r\,(\bar {\bf x}\cdot \bar {\bf dx})}{\bar
r^2+|{\bf a}|^2}-\frac{{\bf a}\cdot ({\bf \bar x}\times d{\bf \bar
x})}{(\bar r^2+|{\bf a}|^2)}-d\bar t\ .\end{equation}Here ${\bf
a}=(c, b, a)$, with $|{\bf a}|^2=a^2+b^2+c^2$ and $\bar {\bf
x}=(\bar x, \bar y, \bar z)$ with $\bar r$ given in terms of $\bar
x, \bar y, \bar z$ by (\ref{1.10}). The dot and cross signify the
usual scalar and vector product in three dimensional Euclidean
space. This form of the Kerr metric has been given by Weinberg
\cite{W}. Weinberg's angular momentum points in the opposite
direction to ours. Now the Riemann tensor is given by
\begin{equation}\label{1.17}
{}^+\bar R_{\mu\nu\rho\sigma}=-\frac{m\,\bar r^3}{(\bar
r^2+i\,({\bf a}\cdot {\bf \bar x}))^3}\,\{\bar
g_{\mu\nu\rho\sigma}+i\,\bar\epsilon _{\mu\nu\rho\sigma}+3\,\bar
W_{\mu\nu}\,\bar W_{\rho\sigma}\}\ .\end{equation} Here $\bar
g_{\mu\nu\rho\sigma}$ has the form (\ref{1.7}) with $\bar
g_{\mu\nu}$ given now by (\ref{1.15}), the permutation symbol is
still $\bar\epsilon _{\mu\nu\rho\sigma}$ on account of its
transformation properties under a general coordinate
transformation (see \cite{SS}) and the fact that the Jacobian of
the transformation (\ref{1.9}) is unity. Also $\bar W_{\mu\nu}$ in
(\ref{1.17}) is now given by
\begin{eqnarray}\label{1.18}
{1\over 2}\bar W_{\mu\nu}\,d\bar x^{\mu}\wedge d\bar
x^{\nu}&=&\frac{\bar x}{\bar r^2+i\,({\bf a}\cdot {\bf \bar
x})}\,[(\bar x+i\,c)\,(d\bar x\wedge d\bar t-i\,d\bar y\wedge
d\bar z)\nonumber\\&&+(\bar y+i\,b)\,(d\bar y\wedge d\bar
t-i\,d\bar z\wedge d\bar x)\nonumber\\&&+(\bar z+i\,a)\,(d\bar
z\wedge d\bar t-i\,d\bar x\wedge d\bar y)]\ .\end{eqnarray}

We make a Lorentz boost in the $\bar x$--direction:
\begin{equation}\label{1.19}
\bar x=\gamma\,(x+v\,t)\ ,\qquad \bar y=y\ ,\qquad \bar z=z
,\qquad \bar t=\gamma\,(t+v\,x)\ ,\end{equation}with $\gamma
=(1-v^2)^{-1/2}$. We are using units in which the speed of light
is unity. As always in such a boost (see \cite{book}, \cite{AS}) the
mass of the source scales as $m=p\,\gamma ^{-1}$ where $p$ is a
constant (the ``energy" of the source). Also the components of the
angular momentum per unit mass orthogonal to the boost direction
(the constants $b$ and $a$) remain finite but the component in the
direction of the boost scales as $c=\hat c\,\gamma ^{-1}$ where
$\hat c$ is a constant (see \cite{PR}, section III for an
explanation of this). The physical origin of this scaling of $c$ 
is that the multipole moments of the isolated source of a gravitational field
experience a Lorentz contraction in the direction of the boost (see 
\cite{multi}). In the case of the Kerr source the multipole moments are
constructed from the mass and the angular momentum per unit mass
(see, for example, \cite{PR} and \cite{Zel}) in such a way that the Lorentz
contraction in the direction of the boost is equivalent to the scaling
of the component of the angular momentum per unit mass $c$ in that direction
given here.
We apply (\ref{1.19}) to the Riemann tensor
(\ref{1.17}) to obtain ${}^+R_{\mu\nu\rho\sigma}$ and then take
the limit $v\rightarrow 1$. The result will be the gravitational
field of the boosted source in the form
\begin{equation}\label{1.20}
{}^+\tilde R_{\mu\nu\rho\sigma}=\lim _{v\rightarrow
1}{}^+R_{\mu\nu\rho\sigma}\ .\end{equation} To evaluate this limit
we shall need generalizations of equations (2.18) and (2.19) in
\cite{PR}. These generalizations are based on the identity
\begin{equation}\label{1.21}
\frac{\bar r^3}{\left (\bar r^2+i\,({\bf a}\cdot {\bf \bar
x})\right )^3}=\frac{1}{[(\bar y+i\,b)^2+(\bar
z+i\,a)^2]}\frac{\partial}{\partial\bar x}\left (\frac{(\bar
x+i\,c)\,\bar r}{\bar r^2+i\,({\bf a}\cdot {\bf\bar x})}\right )\
.\end{equation}Proceeding as in \cite{PR} we derive from this the
limit
\begin{equation}\label{1.22}
\lim _{v\rightarrow 1}\frac{\gamma\,\bar r^3}{\left (\bar
r^2+i\,({\bf a}\cdot {\bf\bar x})\right )^3}=\frac{2\,\delta
(x+t)}{(y+i\,b)^2+(z+i\,a)^2}\ ,\end{equation} where $\delta
(x+t)$ is the Dirac delta function singular on $x=-t$.
Differentiating (\ref{1.22}) with respect to $y=\bar y$ and using
\begin{equation}\label{1.23}
\frac{\partial\bar r}{\partial\bar y}=\frac{\bar r^3\,\bar y+\bar
r\,b\,({\bf a}\cdot {\bf\bar x})}{\bar r^4+i\,({\bf a}\cdot
{\bf\bar x})^2}\ ,\end{equation} we arrive at another useful
limit:
\begin{equation}\label{1.24}
\lim _{v\rightarrow 1}\frac{\gamma\,\bar r^5}{\left (\bar
r^2+i\,({\bf a}\cdot {\bf\bar x})\right
)^5}=\frac{4}{3}\,\frac{\delta (x+t)}{\left
[(y+i\,b)^2+(z+i\,a)^2\right ]^2}\ .\end{equation}

As an illustration of (\ref{1.20}) we find that
\begin{eqnarray}\label{1.25}
{}^+\tilde R_{1212}&=&4\,p\,\left [\frac{z+i\,a+i\,(y+i\,b)}{
(z+i\,a)^2+(y+i\,b)^2}\right ]^2\,\delta(x+t)\nonumber\\&=&4\,p\,\left
[\frac{z+b+i\,(y-a)}{(z+b)^2+(y-a)^2}\right ]^2\,\delta(x+t)\
.\end{eqnarray} This can be rewritten in the form
\begin{equation}\label{1.26}
{}^+\tilde R_{1212}=(h_{yy}-i\,h_{yz})\,\delta (x+t)\
,\end{equation} with
\begin{equation}\label{1.27}
h=2\,p\,\log\{(y-a)^2+(z+b)^2\}\ .\end{equation} Continuing this
process for all components of the Riemann tensor we find that
$\tilde R_{\mu\nu\rho\sigma}\equiv 0$ except for
\begin{eqnarray}\label{1.27'}
{}^+\tilde R_{1212}&=&{}^+\tilde R_{2424}=-{}^+\tilde
R_{1313}=-{}^+\tilde R_{3434}=-{}^+\tilde R_{3134}={}^+\tilde
R_{2124}\nonumber\\&=&(h_{yy}-ih_{yz})\,\delta (x+t)\
,\end{eqnarray}and
\begin{eqnarray}\label{1.28}
{}^+\tilde R_{1213}&=&{}^+\tilde R_{2434}={}^+\tilde
R_{3124}={}^+\tilde
R_{2134}\nonumber\\&=&i(h_{yy}-ih_{yz})\,\delta (x+t)\
,\end{eqnarray}with $h$ given by (\ref{1.27}). This curvature
tensor can be obtained from the metric tensor given via the
line--element \begin{equation}\label{1.29}
ds^2=dx^2+dy^2+dz^2-dt^2-2\,h\,\delta (x+t)\,(dx+dt)^2\
.\end{equation} The process leading to this is described in
\cite{book} pp.96 and 97. This curvature tensor describes an
impulsive gravitational wave with the null hyperplane $x=-t$ as
the history of the wave front. We see from (\ref{1.27}) that this
gravitational field is also singular on $x=-t,\ y=a,\ z=-b$. This
is a null geodesic generator of the null hyperplane $x=-t$.
In the particular case of the boosted Schwarzshild field one
recovers the well--known result \cite{AS} by simply putting
the two angular momentum parameters $b$ and $a$ to zero in (\ref{1.27}).
As noted following eq. (\ref{1.20}) the angular momentum 
per unit mass $c$, in the
direction of the boost, scales differently from the transverse components
$b, a$ in terms of the three--velocity $v$ of the observer, and this
explains why $c$ does not appear in the light--like limit $v \to 1$. 
This shows that the effect of the introduction of the angular momentum
is simply to shift the singularity from $y=z=0$ in the Schwarzschild
case, to $y=a$, $z=-b$ for the Kerr field with angular momentum 
${\bf J}\,=\,(mc,mb,ma)$, the light--like boost being in the $x$--direction. 

The light--like boost of the Kerr gravitational field in an
arbitrary direction has also been studied from a different point
of view in \cite{BN}. In this work the angle $\alpha$ between the
axis of symmetry of the Kerr field and the direction of boost is
introduced and the authors conclude that from their point of
view``only the limiting cases $\alpha =0$ and $\alpha =\pi /2$...
admit a solution in closed form." In addition they point out that
in their analysis ``the general case allows a perturbative
treatment if one suitably rescales the coordinates and expands the
resulting expression with respect to $\sin\alpha$". We have
clearly avoided these restrictions by first tilting the axis of
the Kerr source and then boosting in the $\bar x$--direction. In
addition the positive advantage of our Riemann tensor centered
approach has already been elucidated in \cite{book} and \cite{PR}.

\setcounter{equation}{0}
\section{Scattering Angles}\indent
We consider the deflection of a highly relativistic particle in
the Kerr gravitational field. The direction of the incoming
particle is arbitrary. We call $\bar S$ the rest frame of the Kerr
source and $S$ the rest--frame of the high speed particle. In $S$
the particle sees the Kerr source moving towards it with a speed
close to the speed of light. In the ultrarelativistic limit the
Kerr gravitational field looks to the particle as the
gravitational field of an impulsive gravitational wave. This field
is modelled by the space--time with line--element (\ref{1.29}). In
$S$ the world--line of the particle is a time--like geodesic of
this metric. In $S$ the particle is assumed to be located at
$x=0,\ y=y_0,\ z=z_0$ (say). It starts moving after encountering
the impulsive gravitational wave. Solving the time--like geodesic
equations for the metric given via (\ref{1.29}) we then have in
$S$ the components of the 4--velocity of the particle before and
after encountering the impulsive gravitational wave (see
\cite{CQG}). From these the components of the 4--velocity of the
particle in $\bar S$ are obtained by Lorentz transformation valid
for $v$ close to unity. If $\alpha$ is the angle of deflection of
the high speed particle out of the $\bar x\bar z$--plane after
encountering the gravitational wave and if $\beta$ is the angle of
deflection out of the $\bar x\bar y$--plane then these angles are
given by \cite{CQG}
\begin{equation}\label{2.1}
\tan\alpha = \frac{\dot {\bar y}}{\sqrt{\dot {\bar x}^2+\dot {\bar
z^2}}}=\frac{Y_1}{\{Z_1^2+\gamma ^2[X_1(1-v)+v]^2\}^{1/2}}\
,\end{equation}and
\begin{equation}\label{2.2}
\tan\beta =\frac{\dot {\bar z}}{\dot {\bar
x}}=\frac{Z_1}{\gamma\,[X_1(1-v)+v]}\ .\end{equation} Here the dot
indicates differentiation with respect to proper time and $X_1,
Y_1, Z_1$ are calculated from the function $h(y, z)$ according to
\begin{equation}\label{2.3}
X_1=-\frac{1}{2}[(h_y)_0 ^2+(h_z)_0 ^2]\,,\qquad
Y_1=-(h_y)_0\,,\qquad Z_1=-(h_z)_0\ .\end{equation}The brackets
around a quantity here followed by a subscript zero denote that
the quantity is evaluated at $y=y_0\ , z=z_0$. Also (\ref{2.1})
and (\ref{2.2}) hold for $v$ close to unity. For the particular
function $h$ given by (\ref{1.27}) the angles of deflection
$\alpha$ and $\beta$ are given by
\begin{equation}\label{2.4}\tan\alpha =\frac{-4\,m\,(y_0-a)}{\left [\left
\{(y_0-a)^2+(z_0+b)^2-4\,m^2\right \}^2+16\,m^2(z_0+b)^2\right
]^{1/2}}\ ,\end{equation} and
\begin{equation}\label{2.5}\tan\beta =\frac{-4\,m\,(z_0+b)}{
(y_0-a)^2+(z_0+b)^2-4\,m^2}\ .\end{equation}If the incoming
particle is projected from $y_0=-\eta\ (\eta >0)\ ,\ z_0=0$ then
for large impact parameter $\eta$ these formulas give the small
angles \begin{equation}\label{2.6}\alpha
=\frac{4\,m}{\eta}-\frac{4\,m\,a}{\eta ^2}\ ,\end{equation} and
\begin{equation}\label{2.7}\beta =-\frac{4\,m\,b}{\eta ^2}\ ,
\end{equation} 
approximately. For the reasons given at the end of section 2 the
component of angular momentum in the direction of the incoming 
particle does not appear in the scattering angles.

\setcounter{equation}{0}
\section{Discussion}\indent
The main purpose of this paper is to present the deflection
formulas (\ref{2.6}) and (\ref{2.7}) for the scattering of a high
speed particle in the Kerr gravitational field. When $b=0$ the
formula (\ref{2.6}) agrees with the small angle of deflection of a
photon moving in the equatorial plane of the Kerr source
calculated by Boyer and Lindquist \cite{BL}. This formula has been
calculated in \cite{CQG} and it has been commented on in some
detail there. An important by-product of the present work is the
derivation of the light--like boost of the Kerr gravitational field 
in an arbitrary direction given in section 2.

\noindent
\section*{Acknowledgment}\noindent
One of us (P.A.H.) thanks the C.N.R.S. for a position of Chercheur
Associ\'e.

\end{document}